# Comprehensive Analysis of Dynamic Message Sign Impact on Driver Behavior: A Random Forest Approach

Snehanshu Banerjee, Mansoureh Jeihani, Danny D. Brown, and Samira Ahangari

*Abstract*—This study investigates the potential effect(s) of different Dynamic Message Signs (DMSs) on driver behavior using a full-scale high-fidelity driving simulator. Different DMSs are categorized by their content, structure, and type of messages. A random forest algorithm is used for three separate behavioral analyses – a route diversion analysis, a route choice analysis and a compliance analysis – to identify the potential and relative influences of different DMSs on these aspects of driver behavior. A total of 390 simulation runs are conducted using a sample of 65 participants from diverse socioeconomic backgrounds. Results obtained suggest that DMSs displaying lane closure and delay information with advisory messages are most influential with regards to diversion while color-coded DMSs and DMSs with avoid route advice are the top contributors impacting route choice decisions and DMS compliance. In this first-of-a-kind study, based on the responses to the pre and post simulation surveys as well as results obtained from the analysis of driving-simulation-session data, the authors found that color-blind-friendly, color-coded DMSs are more effective than alphanumeric DMSs - especially in scenarios that demand high compliance from drivers. The increased effectiveness may be attributed to reduced comprehension time and ease with which such DMSs are understood by a greater percentage of road users.

## I. INTRODUCTION

Over the past couple of decades, numerous attempts have been made by researchers to understand the influence of various design, traffic and environmental factors on driving behavior [1-7]. Efforts have also been made to develop accurate model(s) of drivers' behavioral response(s) to these factors [8-12]. The understanding and analysis of driving behavior is vital to the provision of safe driving environments [13-15], as such knowledge is necessary for the development of effective tools designed to influence driver's decisions, speed selection and route choice, and by so doing enhance driving experience and safety. Dynamic message signs (DMS), a component of Advanced Traveler Information System (ATIS), is one such tool deployed to influence driver behavior. Dynamic message signs (DMSs), also known as variable message signs (VMSs) or changeable message signs (CMSs) are readable forms of Intelligent Transportation Systems (ITS) positioned either above or beside a roadway [16], in a manner that is intended to facilitate the efficient and timely transmission of information to road users. They are essential components of numerous ITS and traffic management strategies aimed at regulating, routing or re-routing, warning and managing traffic. Traffic managers use DMSs to influence driver behavior by providing traffic-related information in real time [17]. Figure 1 depicts an example of DMSs displaying information to drivers. A large variety of messages – incident-management, advisory, diversion, special events, adverse road weather condition, speed control, construction, maintenance messages and safety campaign messages – can be displayed on DMSs [18]. These messages are carefully phrased and DMSs strategically positioned to elicit driving behavior that enhances the safety and efficiency of the transportation network as a whole.

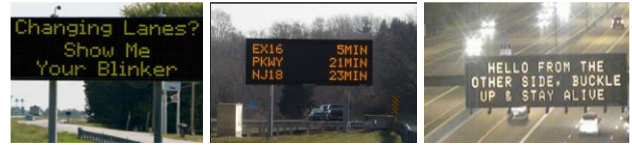

Figure 1: Picture of Dynamic Message Signs (Source: [19])

Pictograms are increasingly being used in DMSs around the world [20]. Tay et al. [20] found that even though most pictograms were easily understood, in cases of incidence occurrence, less than 50% of study participants accurately comprehended information displayed.

The use of DMSs in the United States of America is so widespread that the Department of Transportation (DOT) of 29 states have written guidelines or policies on DMS design and operation [17]. Despite their widespread use, the impact of DMSs on driver behavior and road safety has been questioned and researched by many [4, 21, 22]. Several studies [23-25] show that the type, form, length and phrasing of information presented on DMSs directly affects drivers' level of comprehension. The level of comprehension influences different aspects of driver behavior, especially route choice and compliance. Since diversion and route choice behavior, especially during inclement weather conditions or incident occurrence, are among the top aspects of driver behavior targeted by DMSs, there is a need to study the impact(s) of different message displays on DMS diversion rate as well as driver's compliance and route choice behavior. Very few researchers have utilized a driving simulator to perform DMS-related studies due to a lack of route choice capability in most of the existing driver simulators. A comprehensive view of route choice behavior can be captured only in suitably equipped driving simulators. This becomes possible as alternative routes are revealed when a driver chooses a route and complete information about all possible routes is available. Furthermore, a driving simulator provides a controlled environment with fairly realistic traffic and environmental scenarios, which is not possible in other methods.

This driving simulator-based study fills the aforementioned gap in the literature by analyzing route diversion behavior, route choice and compliance behavior under the influence of different content, structure and type of DMSs using random forest technique. Random forest, a machine learning algorithm, was used for the analysis instead of the predominantly used multinomial logistic or binary logistic regression [26-31] because of its ability to handle complex behavioral classifications that may be nonlinear or involve a multitude of interactions. In addition, random forest has been found to be more accurate than logistic regression as it can handle multicollinearity and presents a different approach to measure important variables [32-35]. It is important to note that although a random forest approach has been used to identify factors that impact crash and injury severity [36-39], the authors found no study that utilized random forest to analyze driver's behavioral response(s) to DMSs or other advanced information systems. In addition to analyzing driver behavior, the study explored the effectiveness of two color-coded/graphic DMS displays, created with consideration for the color blind.

## II. REVIEW OF RELATED WORK

Impacts of DMSs can be studied through different approaches, mainly Stated Preference (SP) and Revealed Preference (RP). The SP approach relies on surveys – mail-back, telephone and, more recently, internet surveys – to gather data regarding mode choice, route choice and other aspects of driver's behavior [40, 41]. Different hypothetical travel scenarios, carefully designed to capture apparent and latent factors that may affect drivers' choice of route, are presented to drivers/participants in a survey. The RP approach, on the other hand, uses data obtained from drivers as they engage in real-world driving scenarios. It comprises of data collected from loop detectors, driver's report, in-vehicle data collection systems and driving simulators.

Levinson and Huo [42] determined the effect of DMSs on total travel time and delay using a before and after study of inductive loop detectors data in different locations in the Twin Cities of Minneapolis and St. Paul, Minnesota. They concluded that even though the DMSs had no significant effect on the reduction of travel time, they were effective in traffic delay alleviation as drivers' diversion increased when warning messages about traffic conditions were displayed. Horowitz et al. [43] reported that even though majority of drivers in their study responded to traffic delay warnings and diverted to the suggested route, a subset of drivers ignored such warning messages and refused to divert. The result was consistent with that obtained by Xuan et. al [44]; some drivers stick to their preplanned route choice regardless of message(s) displayed on DMSs.

Tian et al. [45] categorized the route choice decision making process into strategic and non-strategic and demonstrated with the aid of a driving simulator and surveys that route choice decisions were mainly strategic. However, their result suggested that the cognitive demands of driving affect driver's strategic thinking ability; the simpler the network the more strategic the route choice decision. Jindahra et al [46] used SP data in their quest to determine the route changing propensity of different DMSs in Bangkok, Thailand. They found a correlation between the phrasing of messages on DMSs and diversion rate. Qualitative delay and suggested route information were found to be the most important component of messages intended for route diversion management whereas quantitative messages were important for keeping drivers on the same route. Similar results were obtained by other studies [40, 47] in which questionnaires were used to gather route choice decisions and general behavioral responses of drivers to DMS. Analysis of responses revealed a strong correlation among DMS message content, driver's general behavioral response, the significant effect the information type presented on DMSs, environmental conditions and driver's characteristics have on driver's route choice decisions. Bluetooth sensor technology was innovatively used by Fish et al. [48] in their empirical analysis on the impacts of highway DMSs. The result showed that diversion messages influence the route choice decisions of travelers. Jeihani et al. [49] investigated the factors that influence driver route choice decisions in the presence of DMSs, using a driving simulator integrated with a traffic simulator. They found that the original chosen route, information displayed on DMSs, subject's perception of the relevance, accuracy of displayed information and exposure to DMS affect route diversion. Dia et al. [50] modeled drivers' route choice and compliance behavior using a Neugent model. The model allowed them to capture the dynamic nature of driver behavior and driver's compliance with advisory messages displayed. Simulation results showed that a driver's decision to divert is influenced by how familiar a driver is with network conditions, socio-economic characteristics and the expectation of improvement in travel time by a margin unique to every driver.

## III. METHODOLOGY

A full-scale high-fidelity driving simulator, in the Safety and Behavioral Analysis (SABA) center at Morgan State University, was utilized for this study. The simulator, as seen in Figure 2, is physically comprised of a cockpit, three surrounding monitors to project front and peripheral views as subjects travel though the virtual network, an ignition key, safety seat belt and other components necessary for the operation of the vehicle in the simulated environment: steering wheel, hand brake, throttle, signal light controllers, emergency blinkers, and brake pedals as well as an automatic gear stick.

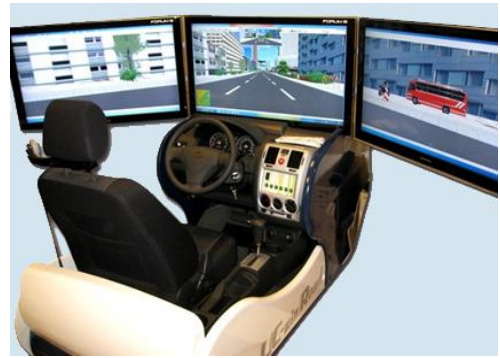

FIGURE 2. Driving Simulator at the SABA Center, Morgan State University

TABLE 1. TYPES OF DMS SIGNS USED IN 6 SCENARIOS

| Order of DMS encountered | Travel time | | Lane Closure | | Delay | |
|---|---|---|---|---|---|---|
| | Scenario 1 | Scenario 2 | Scenario 3 | Scenario 4 | Scenario 5 | Scenario 6 |
| DMS-1 | Distance time | Distance time | Crash related | Crash | Color Coded | Color Coded |
| | With alternative routes | W/O alternative routes | With avoid advice | W/O advice | Design I | Design II |
| DMS-2 | Distance time | Travel time | Lane closure | DMS | DMS | Delay |
| | With alternative routes | With alternative routes | With alternate route | With avoid advice | With save time advice | With advice |
| DMS-3 | Travel time | Travel time | Crash | DMS | Delay | Delay |
| | With alternative routes | W/O alternative routes | With advice | W/O advice | With advice | W/O advice |
| DMS-4 | Distance time | Travel time | Lane closure | Incident | N/A | N/A |
| | W/O alternative routes | W/O alternative routes | W/O advice | W/O advice | | |

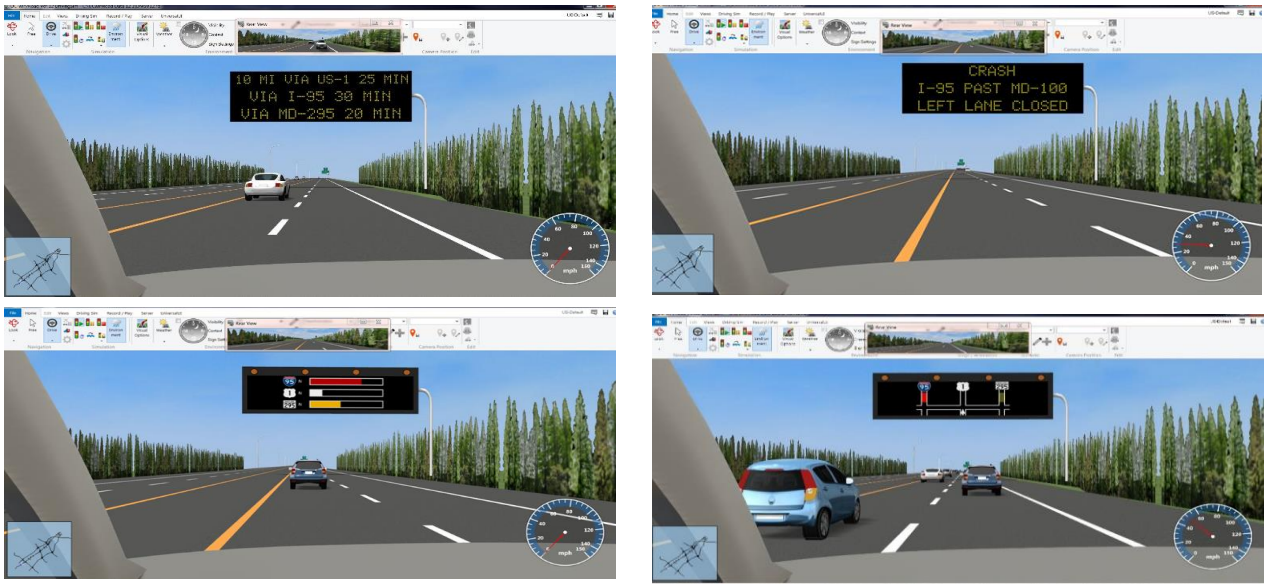

FIGURE 3. A screenshot of the Simulated Driving Environment and some DMS signs

The use of a driving simulator enables researchers to capture the effect(s) of environmental factors and surrounding traffic on subjects' compliance, diversion and route choice decision(s) –an essential component missing from SP data collection methods.

Human subjects, henceforth referred to as participants, were asked to drive from a clearly defined origin through the virtual road network to a fixed destination. Participants were free to choose and change routes as they drove through 6 different scenarios consisting of different DMS contents, types, structure, and length.

*A. Scenario Design*

Six virtual driving scenarios were created, with the aid of a proprietary software-VR-design studio-developed by FORUM Co [51]. The types of DMSs are shown in Table 1. DMS-4 from all scenarios weren't used in this study as they were near the destination and the behavior in terms of route choice, diversion or compliance and could not be recorded for lack of an exit route. The virtual scenarios were complete with traffic lights, trees, building structures and other objects as seen in Figure 3. Driving behavior data: brake, throttling and steering handling parameters, as well as route choice data were automatically recorded by the driving simulator. However, for this study, only route information was utilized.

*B. Network Characteristics*

A study area of 400 square kilometers (20x20 km) southwest of the Baltimore metropolitan area was utilized for this study. The origin was set at the Washington Blvd-Montevideo intersection while the destination was fixed at the M&T Bank stadium (intersection of Russell Street and Baltimore-Washington Parkway (MD-295)). Google Maps was used as the reference to develop all roadway signs, trees and intersections in the virtual network similar to the real world. Realism in simulation sessions was achieved by carefully setting traffic volume and characteristics to emulate those obtainable in real-life driving environments. Figure 4 shows the study network, the origin and destination of the study, and the location of DMSs. As seen from the Figure 4, the network has nine decision points (at which participants can switch routes between US-1, I-95 and I-295) and 10 DMSs locations, four of which are on US-1, three on I-295 and three on I-95. As presented in Figure 4, there are three routes between the origin and the destination. I-95 is a 4-lane interstate route with a speed limit of 65 mph in the study area. Washington Blvd (US-1) is a 2-lane highway with a speed limit of 40 mph. The study area has frequent traffic signals on US-1. MD-295 is a 2-lane highway which expands into a 3-lane, with a speed limit of 55 mph. During non-peak hours, I-95 is typically the fastest route, taking between 12 – 16 minutes to reach M&T Bank stadium from the origin in the study area whereas MD-295 and US-1 take anywhere between 14 – 18 minutes and 14 – 20 minutes respectively. In the study, traffic on I-95 was designed to be heavy to test behavior of drivers who are acquainted with I-95. Traffic on US-1 and MD-295 were set to mimic real-life non-peak hour traffic. In this study, the 3 routes are connected via MD-100, I-195 and I-695 respectively. A categorical list of some of the DMSs displayed in different scenarios of the study as mentioned in Table 1 is shown in Table 2.

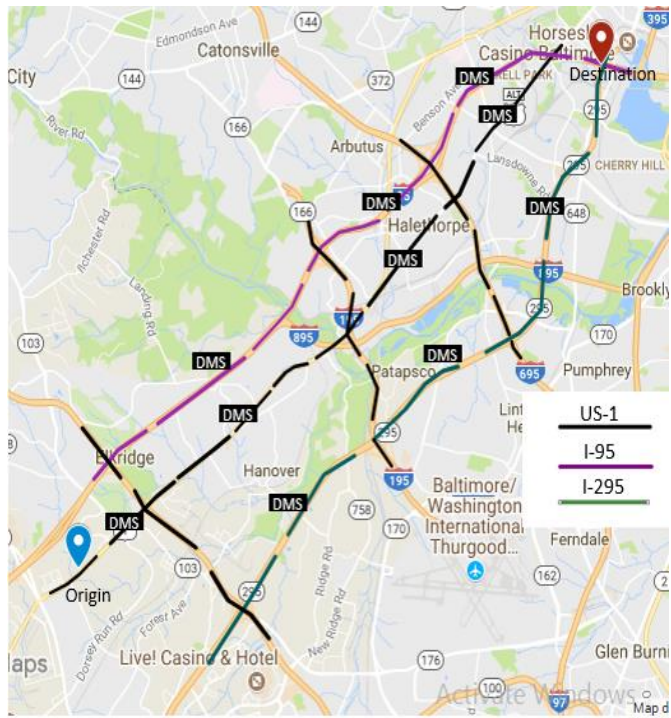

Figure 4. Study Network

TABLE 2. LIST OF DMS SIGNS UTILIZED IN THIS STUDY

| DMS categories | Signs Used |
| --- | --- |
| Distance Time with Alternate Routes | 5 MI VIA I-95 12 MIN / VIA US-1 8 MIN / VIA MD-295 5 MIN |
| Travel Time with Alternate Routes | STADIUM 28 MIN / VIA US-1 15 MIN / VIA MD-295 12 MIN |
| Travel Time Without Alternate Routes | STADIUM 12 MIN |
| Lane Closure Information with Alternate Route | ROADWORK PAST I-195 / LEFT LN CLOSED / USE MD-295 |
| Crash Related DMS With Advice | CRASH PAST I-695 / CONSIDER ALT ROUTE |
| Delay Related DMS With Advice | CRASH AHEAD 1 MI / 15 MINUTES DELAY / USE MD-295 |
| Delay Related DMS Without Advice | CRASH AHEAD 1 MI / 15 MIN DELAY |
| Color-Coded DMS (Design II) | (color-coded route status graphic) |

| DMS With Avoid Route Advice | 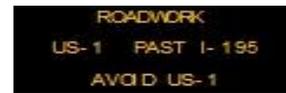 ROADWORK US-1 PAST I-195 AVOID US-1 |
| --- | --- |
| DMS With Save Time Advice | 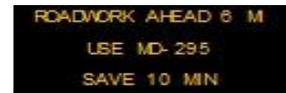 ROADWORK AHEAD 6 M USE MD-295 SAVE 10 MIN |

*C. Survey Questionnaires*

Eight surveys – two pre-simulation and six post-simulation surveys – were designed to capture essential information about participants. Of the pre-simulation surveys, the first survey captured participants' gender, age, household income, educational status and other socio-economic data as well as participants' familiarity with and trust in messages displayed in DMSs and compliance. The second pre-simulation survey attempted to determine participants' familiarity with the study area, their preferred route, level of comprehension of messages displayed on DMSs and level of preference of different types of messages. A post simulation survey was completed by the participant after each of the six scenarios to test the participant's comprehension of displayed messages and recollection of DMSs encountered.

*D. Recruitment process*

Institutional Review Board (IRB) approval was received before human participants were recruited. Social media advertisement, word of mouth and fliers were utilized to recruit participants to drive the simulator. Participants were compensated at the rate of $15 per hour for their involvement in the study. A total of 68 participants were recruited but only 65 completed all scenarios. A total of 390 simulation sessions were conducted and recorded. Participants drove for five to 10 minutes to become familiar with the driving simulator environment before driving the six scenarios. They were also given a 5-minute break between scenarios to avoid fatigue. Rules were set to ensure participants handled the simulator as they would their vehicle in the real world. Warnings, red-light running and speeding tickets, in the form of deduction(s) from

compensation/payments, were randomly issued for non-compliance with traffic rules and for crashes to ensure driving realism.

## IV. DATA

In this study, the data collected from the surveys, sociodemographics of the participants and category of DMS signs were used as predictor variables. Diversion, compliance and route choice were the response variables in the three separate datasets, created for behavioral analysis. In a bid to determine the impact of DMS messages on driver behavior, all the categories of message types were transformed to separate dummy variables. The datasets were unbalanced due to drivers' route choices, with some signs being less frequently encountered. Although a random forest algorithm handles categorical data well, it is biased towards categorical variables with a high number of levels [52]. To address this issue, all the categorical variables were converted to dummy variables to improve outcome interpretability. Descriptive statistics of the sociodemographic and survey data after this transformation, used in all 3 datasets, are shown in Table 3.

TABLE 3. SOCIO-DEMOGRAPHIC AND SURVEY DATA DESCRIPTIVE STATISTICS

| Variables | Description | Percentage |
|---|---|---|
| Gender | Male | 55% |
| | Female | 45% |
| Age | 18 – 25 | 33% |
| | 26 – 35 | 39% |
| | 36 – 45 | 11% |
| | 46 – 55 | 10% |
| | 56 - 65 | 7% |
| Familiarity with Study Area | Yes | 53% |
| | Somewhat | 28% |
| | No | 13% |
| Frequency of Travel | Very frequently | 25% |
| | Often | 37% |
| | Occasionally | 24% |
| | Never been there | 9% |
| Route Usually Taken | MD-295 | 19% |
| | US-1 | 5% |
| | I-95 | 34% |
| | Follow my GPS | 30% |
| | Not Sure | 8% |
| DMS Influences Decisions | Always | 18% |
| | Sometimes | 77% |
| | Never | 3% |
| When DMS GPS Conflict | I follow DMS | 27% |
| | I follow GPS | 38% |

The sum of the percentages for some variables shown in Table 3 may not add up to 100%, as some sections in the survey were left blank by the participants. Analyses were carried out using the open source R-project statistical software [53].

### A. Random Forest

Random forest is a supervised learning algorithm which can be used for both classification and regression modeling [54]. It consists of an ensemble of decision trees, i.e. CART (classification and regression trees), commonly trained with the bagging technique where the idea is to combine multiple models to improve classification accuracy thereby reducing the risk of overfitting [55]. The decision trees in a random forest are grown from bagged samples from the training set using the following equation [34]:

$$Y = \sum_{j=1}^{r} \beta_j I(x \in R_j) + \varepsilon$$

where the coefficients $\beta_j$ is the average of the Y values in the region $R_j$ which is estimated from the dataset. Once the set of decision trees have grown, the unsampled observations are dropped down each tree from the test dataset and these 'out of bag' (OOB) observations are used for internal cross validation and to calculate prediction error rates. The error calculated is the mean decrease in node impurity (mean decrease Gini or MDG) which can be used for variable selection by ranking variables in the order of importance. The random forest package in "R" [56] was used to compute MDG which is the sum of all decreases in Gini impurity due to a given variable and then normalized toward the end of the forest growing stage. MDG is the predictive accuracy lost by permuting a given predictor variable from the tree used to generate predictions about the class of observation $i$, where $i \in [0,1]$, the Gini score range. Thus, predictor variables with a higher MDG score more accurately predict the true class of observation $i$ which is also termed as the variable importance measure (VIM) in random forests. As VIMs are not sufficient to capture the trend of influence of the predictor variables on the dependent variable, partial dependency plots (PDP) are used to address this limitation. PDPs offer a graphical portrayal of the marginal effect of a variable on the class probability. The function is represented mathematically as [34]:

$$f(x) = \log p_n(x) - \frac{1}{N} \sum_{m=1}^{n} \log p_m(x)$$

where N represents the number of classes for the dependent variable $y$, $n$ is the predicted class and $p_m$ represents the fraction of votes for class m [57]. On a PDP, the values of the $y$ axis indicate the change in log-odds for the fraction of votes among all decision trees in the forest. A positive slope indicates that the variable predicts a greater fraction of votes for that particular class of the dependent variable and vice versa.

V. RESULTS

The following sub-sections discuss in detail, the findings of this study.

*A. Stated Preference Vs Revealed travel behavior*

In response to the survey, most participants indicated a preference for either I-95 or whatever route was suggested by a global positioning system (GPS) or a smartphone to go from the origin to the destination in the study area. The stated route choice of participants as obtained from the survey responses is shown in Figure 5. However, disparity was observed between route choice selected in survey and route choice during the simulation sessions. This suggests that route choice was influenced by DMSs and environmental conditions. Figure 6 displays the categories of DMSs that potentially influenced route choice decisions.

"Distance time with alternate routes" and "Distance time" DMSs don't seem that effective in influencing a particular route as shown in Figure 6. "Color-coded DMS" and "Crash-related DMS," on the other hand, seem effective in influencing route choice decisions. Design II was the preferred option in comparison to Design I of color coded DMS as mentioned in the SP survey.

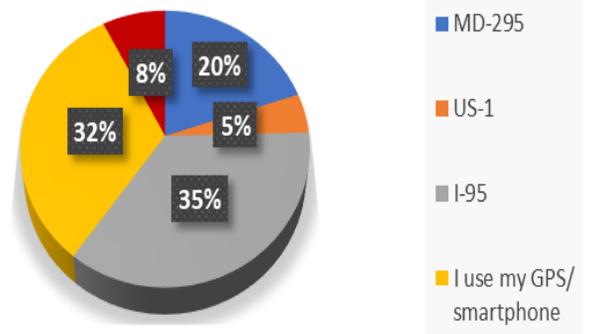

Figure 5. Stated route choice responses

*B. Route diversion behavioral analysis*

For this analysis, only DMSs preceding an exit ramp were selected, from the 6 scenarios, to examine the patterns of diversion in response to messages displayed on such DMSs. In all scenarios, the first DMS encountered by participants was excluded from this behavioral analysis to avoid biases that may arise from a participant's pre-selected choice of route. With the aid of the random forest algorithm, the sociodemographic and survey data as mentioned in Table 3 and the DMS categories mentioned in Table 4 were used for this analysis.

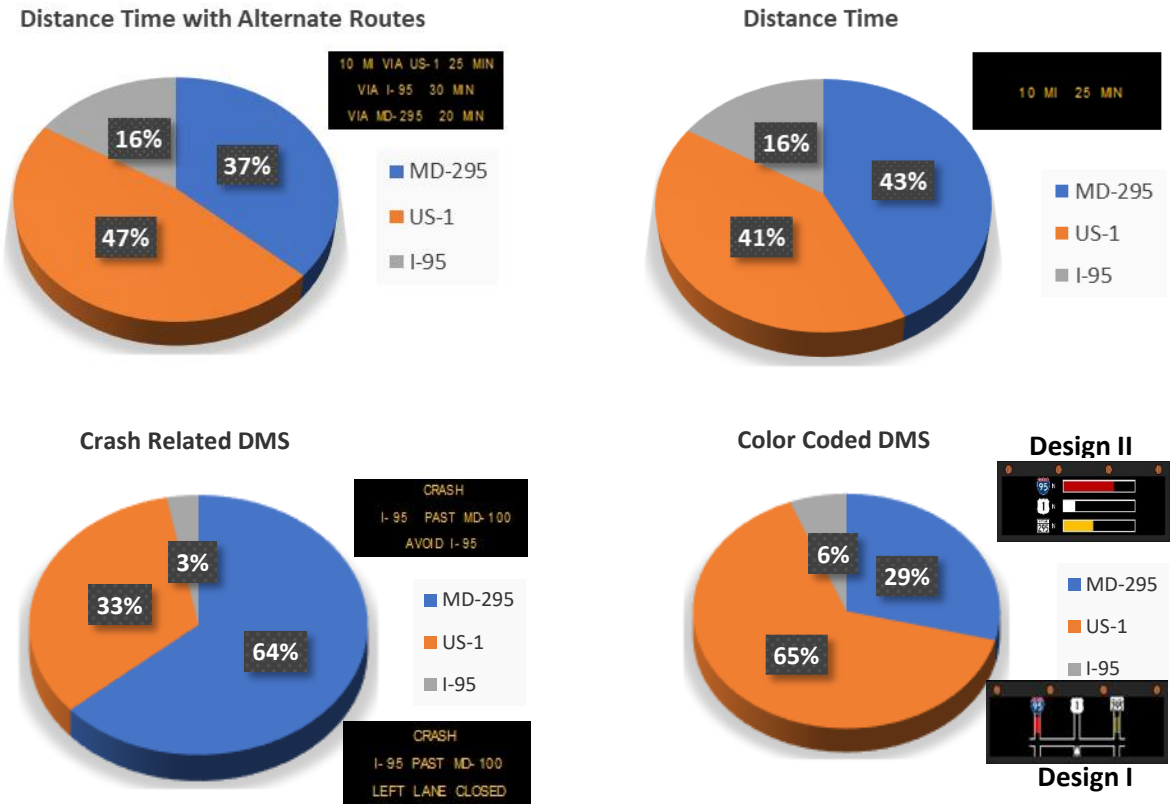

Figure 6. Revealed route choice behavior

TABLE 4. DESCRIPTIVE STATISTICS OF ROUTE DIVERSION DATASET

| Variables | Description | Percentage |
|---|---|---|
| Distance Time with Alternate Routes | Encountered | 11% |
| | Did not encounter | 89% |
| Travel Time with Alternate Routes | Encountered | 22% |
| | Did not encounter | 78% |
| Travel Time without Alternate Routes | Encountered | 12% |
| | Did not encounter | 88% |
| Lane Closure Information with Alternate Route | Encountered | 11% |
| | Did not encounter | 89% |
| Crash Related DMS With Advice | Encountered | 11% |
| | Did not encounter | 89% |
| Delay Related DMS With Advice | Encountered | 22% |
| | Did not encounter | 78% |
| Delay Related DMS Without Advice | Encountered | 11% |
| | Did not encounter | 89% |
| Diversion | Diverted | 42% |
| | Did not divert | 58% |

The OOB classification error estimate for this analysis was 36.14% built with 500 trees in the forest. Figure 7 shows the MDG score for all the variables used for route diversion analysis. It can be seen that 4 variables (Travel time without alternate routes, lane closure information with alternate routes, delay-related DMS with advice and when

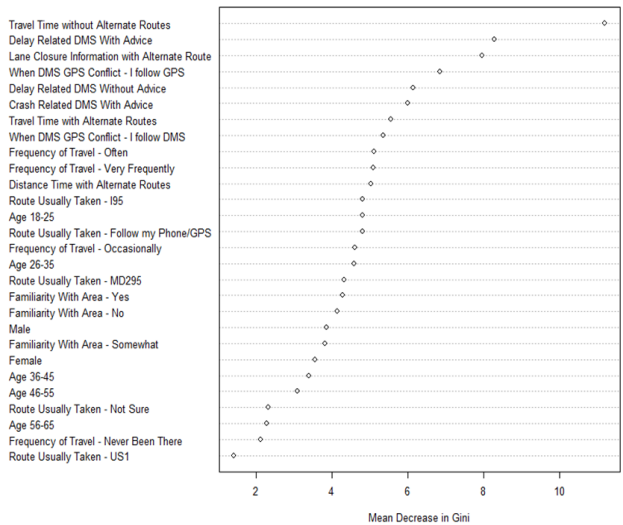

Figure 7. Plot of variable importance for diversion by MDG score

DMS/GPS conflict – I follow GPS) stand out and are thus selected as the important variables. To determine the trend of influence these variables have on diversion, PDPs were drawn as shown in Figure 8. The PDPs for this dataset are bar charts with binary outcomes with an increasing or decreasing trend as shown in Figure 8.

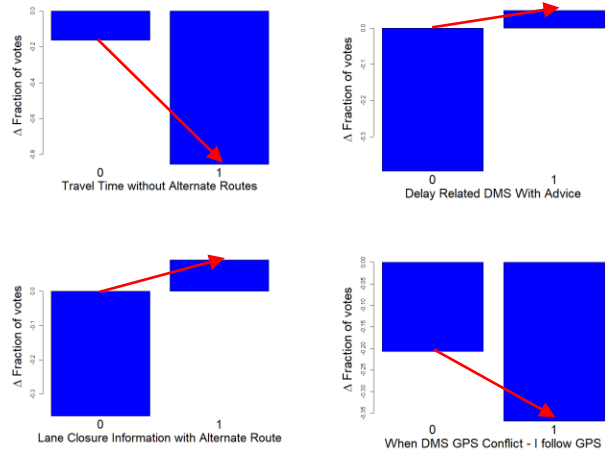

Figure 8. PDPs of important variables impacting diversion behavior[1]

Although travel time without alternate route is the most important variable, it has a negative influence on diversion which means that it is ineffective for diversion. This might be due to the non-provision of any pertinent information related to diversion other than just the travel time to the destination. Drivers who stated that they would follow their GPS in case the DMS message conflicts with their GPS were less likely to divert from the chosen route even in the absence of a navigation system. Delay-related messages with advice and lane closure messages with alternate route information were found to have a positive influence on diversion. This means that DMSs displaying such messages will most likely impact route diversion decisions.

*C. Route choice behavioral analysis*

For this analysis, the first sign participants encountered in the network was selected to determine route choice behavior.

TABLE 5. DESCRIPTIVE STATISTICS OF ROUTE CHOICE DATASET

| Variables | Description | Percentage |
|---|---|---|
| DMS Messages | Distance Time with Alternate Routes | 17% |
| | Distance Time | 17% |
| | Crash Related DMS with Advice | 33% |
| | Color Coded DMS | 33% |
| Route Choice | MD-295 | 44% |
| | US-1 | 47% |
| | I-95 | 9% |

---

[1] The direction of the trend is shown by the red arrows.

The sociodemographic and survey data as mentioned in Table 3 and the sign categories mentioned in Table 5 were used for this behavioral analysis.

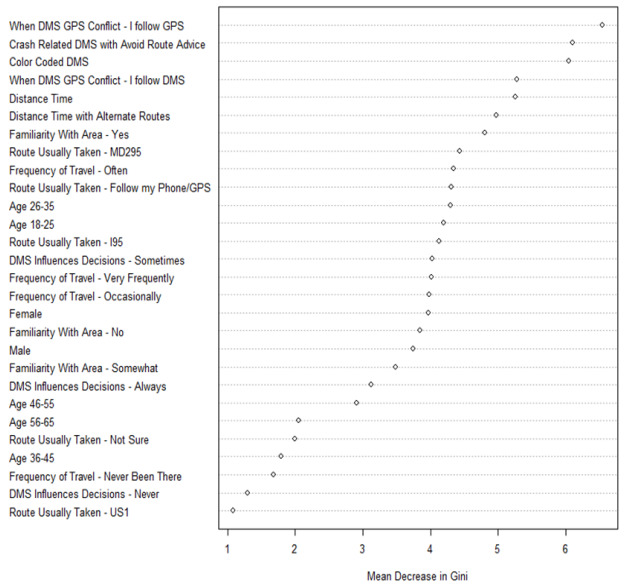

Figure 9. Plot of variable importance for route choice by MDG score

The OOB error estimate for this analysis was 39.86% built with 500 trees in the forest. Figure 9 shows the MDG score for all the variables used for route choice analysis. The results show that 3 variables (color-coded DMS, crash-related DMS and when DMS/GPS conflict – I follow GPS) stand out and are thus selected as the important variables.

To determine the trend of influence these variables have on route choice, PDPs were drawn for each class, as shown in Figure 10. The binary outcomes for each route are shown for the selected important variables.

Color-coded DMS was found to be the most important variable and, as can be seen in Figure 10, participants have a higher likelihood to pick US-1 over other routes as the DMS showed heavy traffic on I-95, medium traffic on MD-295 and light traffic on US-1. Similarly, the crash-related DMS with avoid route advice rendered information of a crash on I-95 and advised participants to avoid it. Participants responded to the DMS by using either MD-295 or US-1. Participants who answered that they would follow GPS in case of conflicting DMS route suggestions showed less likelihood of picking I-95 in the absence of a GPS, as advised by the DMS to avoid that route.

### D. DMS compliance behavioral analysis

All signs with advisory messages were selected to test compliance. The sociodemographic and survey data as mentioned in Table 3 and the sign categories mentioned in Table 6 are used for this behavioral analysis.

The OOB error estimate for this analysis was 36.62% built with 500 trees in the forest. Figure 11 shows the MDG score for all the variables used for DMS compliance analysis. The results show that 4 variables (Distance time with alternate routes, color-coded DMS, DMS with avoid route advice and crash-related DMS with advice) stand out and are thus selected as the important variables. To determine the trend of influence these variables have on message compliance, PDPs were drawn as shown in Figure 12.

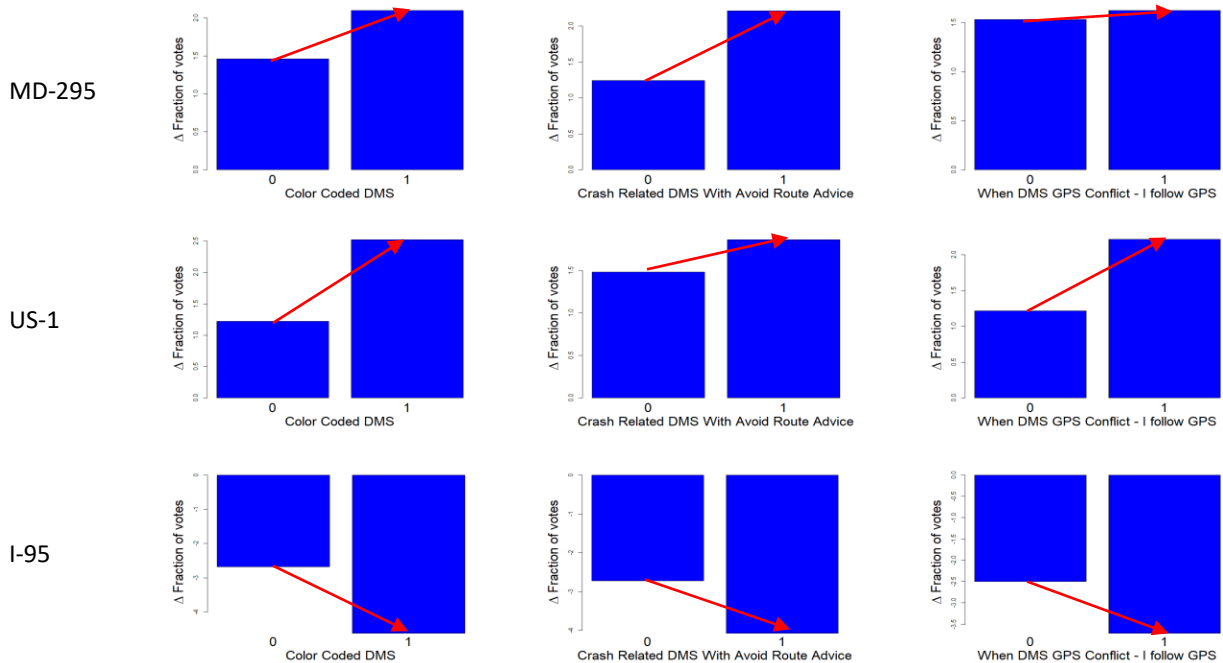

Figure 10. PDPs of important variables impacting route choice[2]

---

[2] The direction of the trend is shown by the red arrows.

TABLE 6. DESCRIPTIVE STATISTICS OF COMPLIANCE DATASET

| Variables | Description | Percentage |
|---|---|---|
| DMS Messages | Distance Time with Alternate Routes | 18% |
| | Travel Time with Alternate Routes | 18% |
| | Color Coded DMS | 17% |
| | Lane Closure Information with Alternate Routes | 9% |
| | Crash Related DMS With Advice | 8% |
| | DMS With Avoid Route Advice | 4% |
| | Delay Related DMS With Advice | 17% |
| | DMS With Save Time Advice | 9% |
| Compliance | Complied | 53% |
| | Did not comply | 47% |

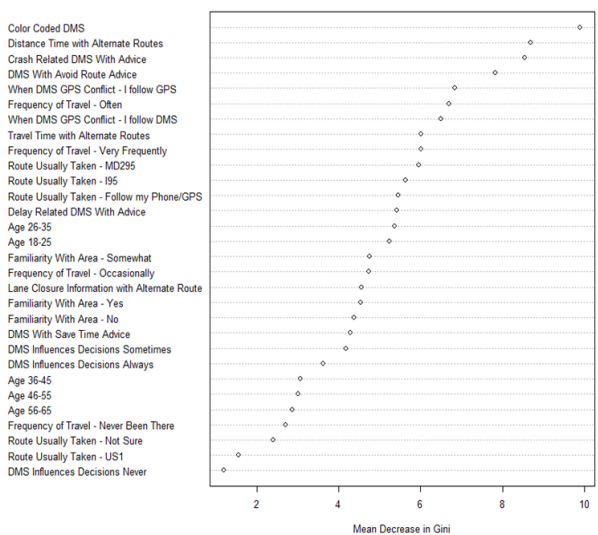

Figure 11. Plot of variable importance for compliance by MDG score

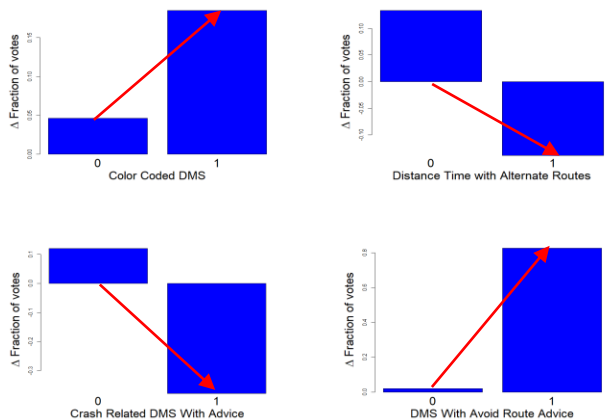

Figure 12. PDPs of important variables impacting compliance behavior

Although DMS with 'distance time with alternate routes' message is an important variable, it has a negative influence on compliance. This may be attributed to the very low travel time differences between the 3 routes, a maximum difference of 10 minutes between all of them. Results from the compliance analysis also showed that "Crash-related DMS with advice," had a high likelihood of non-compliance. This might be because the advice tested in this study was vague, "choose alternate route." Color-coded DMS had a higher likelihood of compliance as the color-coded DMS was easy to comprehend (as expressed in survey responses) and time taken to perceive it is less than alphanumeric text [20]. Avoid route advice, on the other hand, is very specific and is most likely the reason why the compliance rate is high.

## VI. CONCLUSION AND DISCUSSION

This study investigated the impact of content, structure and type of DMS messages on driver behavior using a full-scale high-fidelity driving simulator. A route diversion analysis, a route choice analysis and a DMS compliance analysis were conducted using a random forest algorithm to gauge how people react to different signs and how it impacts their decision making. Prior research on this study corridor showed that people have a tendency of usually choosing I-95 as their default route, since it is wider, has a higher speed limit and is faster under normal traffic conditions [58]. This driving behavior is altered under non-recurrent situations like roadwork or crashes. In this study, the DMS messages have stated throughout the 6 scenarios that I-95 had heavy traffic. The results indicate that the participants tend to comply with crash-related DMSs with advice, especially advice which mentions "avoid," lane closure with alternate route advice and delay-related DMS with advice. In contemporary times, the majority of drivers depend on their GPS/smartphones for turn-by-turn guidance to reach their destination. Some 98% of the participants in this study stated that they use GPS/smartphone for navigation at least sometimes. In such scenarios, drivers hardly pay attention to travel time-related DMS messages. Smartphone navigation informs the driver of a delay before even starting from the point of origin. But incidents like crashes can happen at any time and appropriate DMS messages are useful in such situations, which can prevent delay and congestion. Lane closure, delay DMS with route diversion and avoid route information will likely be useful in such situations once drivers start experiencing a slow down on their choice of route.[3]

One of the more interesting findings in this study is the effectiveness of color-coded DMS on driver compliance and route choice. What makes it interesting is the fact that input from color blind people was used to design the DMS. Approximately 7.5% (~11 million) of the United States population cannot distinguish between red and/or green colors. As they have become familiar with traffic lights, the authors designed the color coded DMS messages to be color blind people-friendly. Although the red and yellow colors on the DMS sign were to show heavy traffic and medium traffic, they were in the shape of horizontal bars as shown in Figure 6 (Design II). The length of the bars would depict traffic congestion levels making them colorblind people-friendly based on the input received. Even though there weren't any color-blind participants, the overall compliance of color-coded DMS, and its effectiveness in determining route choice, makes it a valuable means of signage. DMS displays in the United States are predominantly alphanumeric text as opposed to graphical displays [59]. The New Jersey turnpike and some freeways in California employ color-coded DMS to direct traffic. Color-coded DMS has not yet found widespread use in the United States, and its effectiveness has yet to be tapped. Pilot studies can be carried out in certain corridors using color-coded DMSs to balance traffic flow on congested routes.

---
[3] The direction of the trend is shown by the red arrows.

One of the limitations of this study was the lack of a navigation system to supplement the DMS messages. Future studies will involve scenarios with GPS guidance as well as interaction with DMS messages. Another aspect that this study didn't touch on, was from a DMS security standpoint. If message signs are hacked, that can easily influence driver behavior and that can create issues, causing congestion as well as crashes. Several studies have been conducted on this topic [60-63]; future studies will consider integrating DMSs from a security standpoint with driver behavior.

ACKNOWLEDGMENT

The authors would like to thank the Maryland State Highway Administration for its funding support throughout the study.